\documentclass[
 reprint,
 amsmath,amssymb,
 aps,
 floatfix,
]{revtex4-2}

\usepackage{graphicx}
\usepackage{bm}
\usepackage{hyperref}

\begin{document}

\title{Anomalous First Passage in Evolution: Edge-KPZ Theory}
\author{Tetsuhiro S. Hatakeyama}
\email{hatakeyama@elsi.jp}
\affiliation{Earth-Life Science Institute, Institute of Future Science, Institute of Science Tokyo, 2-12-1-IE-1 Ookayama, Meguro-ku, Tokyo 152-8550, Japan}
\date{September 2026}

\begin{abstract}
The pace of evolution depends on how rapidly new phenotypes arise.
We show that neutral Wright--Fisher evolution exhibits anomalous first
passage despite diffusive mutations.
The mean time for the first individual to reach a prescribed phenotypic
distance scales approximately as $(\sigma^2)^{-3/2}$ with mutation
variance $\sigma^2$.
Two crossovers bound this regime, with inverse-variance scaling on either
side.
Combining coalescent theory with Kardar--Parisi--Zhang (KPZ) fluctuations
at the dilute population edge, we develop an edge-KPZ theory of all three
regimes.
The anomaly persists under weak selection.
\end{abstract}

\maketitle

\section{Introduction}

The pace of evolution has long been a central question in evolutionary
biology and biophysics.
This pace depends in part on the accumulation of mutations in regions
where selection is absent or weak \cite{Kimura1968,Ohta1973}.
As populations explore different genetic backgrounds while retaining
existing functions, they can gain access to new phenotypes and adapt more
readily to new conditions \cite{Wagner2008,Hayden2011}.
Despite the extensive theory of fitness-valley crossing
\cite{Iwasa2004,Weissman2009}, a systematic understanding of how mutation
and reproduction jointly determine the time to reach a distant phenotype
remains incomplete, even in the simplest case of neutral evolution.

The need for such a theory may have been overlooked because neutral
exploration appears to be an ordinary diffusion problem.
Independent, unbiased mutations make each individual's
phenotype perform a random walk along its ancestral lineage.
Writing the mutation variance per generation as
$\sigma^2$, the corresponding diffusion coefficient in phenotype space is
$\sigma^2/2$.
The first-passage time measures how long it takes for the first individual
in a population to reach a prescribed distance from the initial phenotype
in either direction.
For a single diffusing individual, its mean is inversely proportional to
the diffusion coefficient \cite{Gardiner2009} and hence scales as
$(\sigma^2)^{-1}$.
Reproduction connects these walks through shared ancestry, but the
population may still seem to be simply a collection of diffusing
individuals. More individuals would then speed up first arrival while
preserving this inverse-variance law.

Here we show that neutral Wright--Fisher evolution violates this
diffusive expectation: the mean first-passage time scales approximately
as $(\sigma^2)^{-3/2}$ over an intermediate range of mutation variances,
with population size and target distance held fixed. Two crossovers bound
this anomalous regime, with the usual inverse-variance dependence
recovered on either side.
We combine Kardar--Parisi--Zhang (KPZ)-type growth fluctuations with
coalescent genealogy and a microscopic density description in the spirit
of Dean--Kawasaki theory to formulate an edge-KPZ theory.
It connects fluctuations at the dilute population
edge to the anomalous scaling and links its amplitude to the two
crossover scales. The anomalous regime also persists under weak selection.

\section{Model}
\label{sec:model}

The Wright--Fisher process models reproduction as random sampling of a
finite population from one generation to the next
\cite{Fisher1930,Wright1931}.
In the asexual version considered here, a population of $N$ individuals
is replaced each generation by $N$ offspring.
Each offspring independently chooses a parent with probability
proportional to that parent's reproductive fitness and inherits its
phenotype, modified by mutation.
An individual can thus leave several offspring or none, while the total
population size stays fixed. The fitness weights represent selection;
random fluctuations in offspring numbers give rise to genetic drift.

We first consider the simplest case, neutral evolution, in which all
phenotypes have the same reproductive fitness. Every individual is then
equally likely to be chosen as a parent, with probability $1/N$.
Let $X_1(t),\ldots,X_N(t)$ denote the one-dimensional phenotypes at
generation $t$, initially $X_i(0)=0$.
Writing the parent chosen by offspring $i$ as $A_i(t)$, we take mutations
to be independent Gaussian displacements:
\begin{equation}
    X_i(t+1)=X_{A_i(t)}(t)+\sigma Z_i(t),
    \qquad Z_i(t)\sim\mathcal N(0,1).
    \label{eq:wf_update}
\end{equation}
The parameter $\sigma^2$ is the mutation variance per
individual per generation; the corresponding continuum diffusion
coefficient is $\sigma^2/2$.

Our observable is the first-passage time (FPT) for any individual to exit $[-L,L]$:
\begin{equation}
    T_L=\inf\left\{t\geq1:\max_{1\leq i\leq N}|X_i(t)|>L\right\}.
    \label{eq:fpt_definition}
\end{equation}
The threshold represents a prescribed phenotypic displacement. Crossing
it measures the first production of an extreme phenotype; establishment
and fixation would be subsequent evolutionary events.
Related first-passage observables have been used to study polyploid
evolution~\cite{HatakeyamaOhbayashi2022}.

\begin{figure*}[tb]
    \centering
    \includegraphics[width=0.88\linewidth]{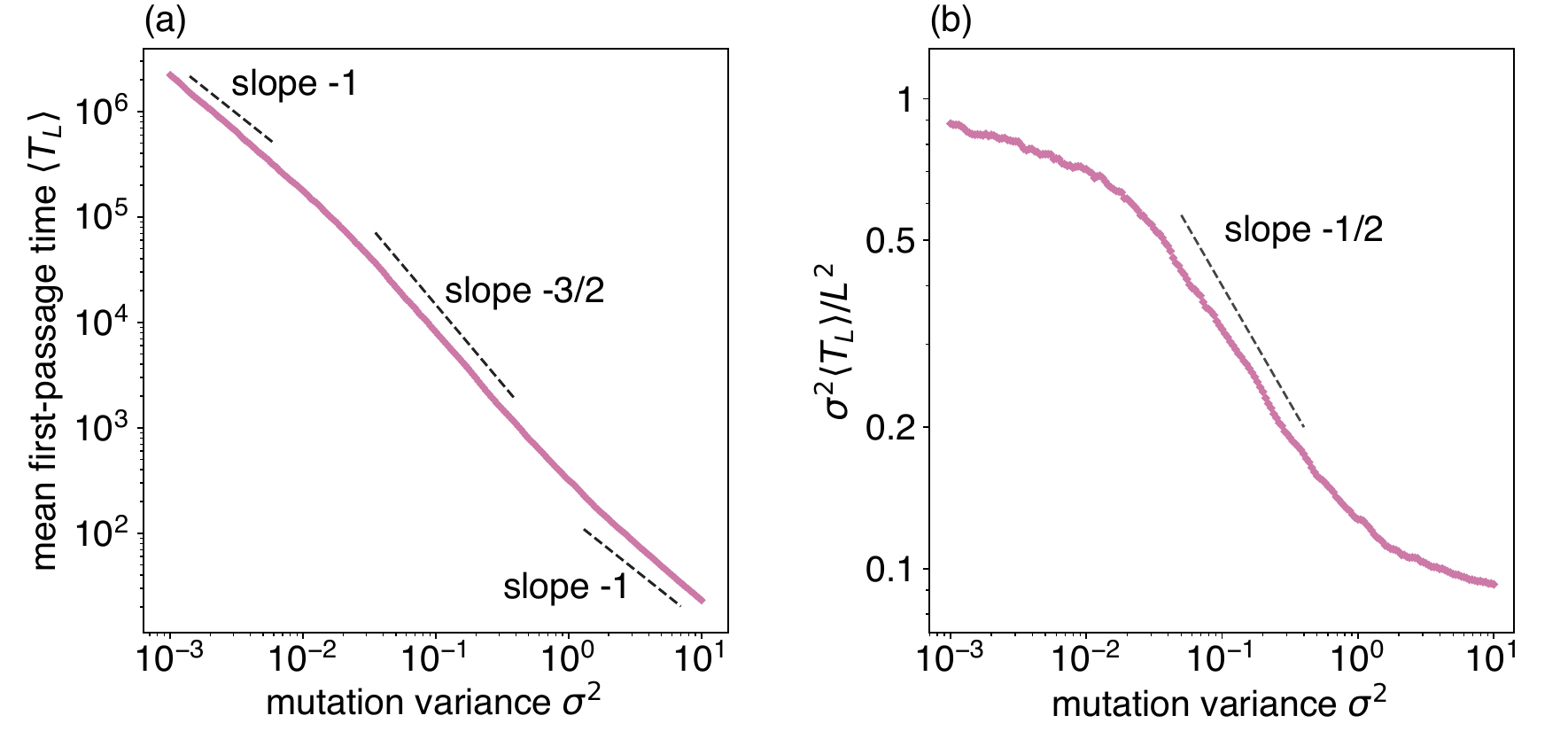}
    \caption{Mean first-passage times in neutral evolution.
    (a) Mean first-passage time for any individual to exit $[-L,L]$ at $N=1000$ and $L=50$,
    as a function of the mutation variance $\sigma^2$.
    (b) The same data multiplied by $\sigma^2/L^2$.
    Symbols represent means over 1000 realizations at each parameter
    point. Dashed segments are slope guides.}
    \label{fig:discovery_sequence}
\end{figure*}

\begin{figure*}[tb]
    \centering
    \includegraphics[width=0.88\linewidth]{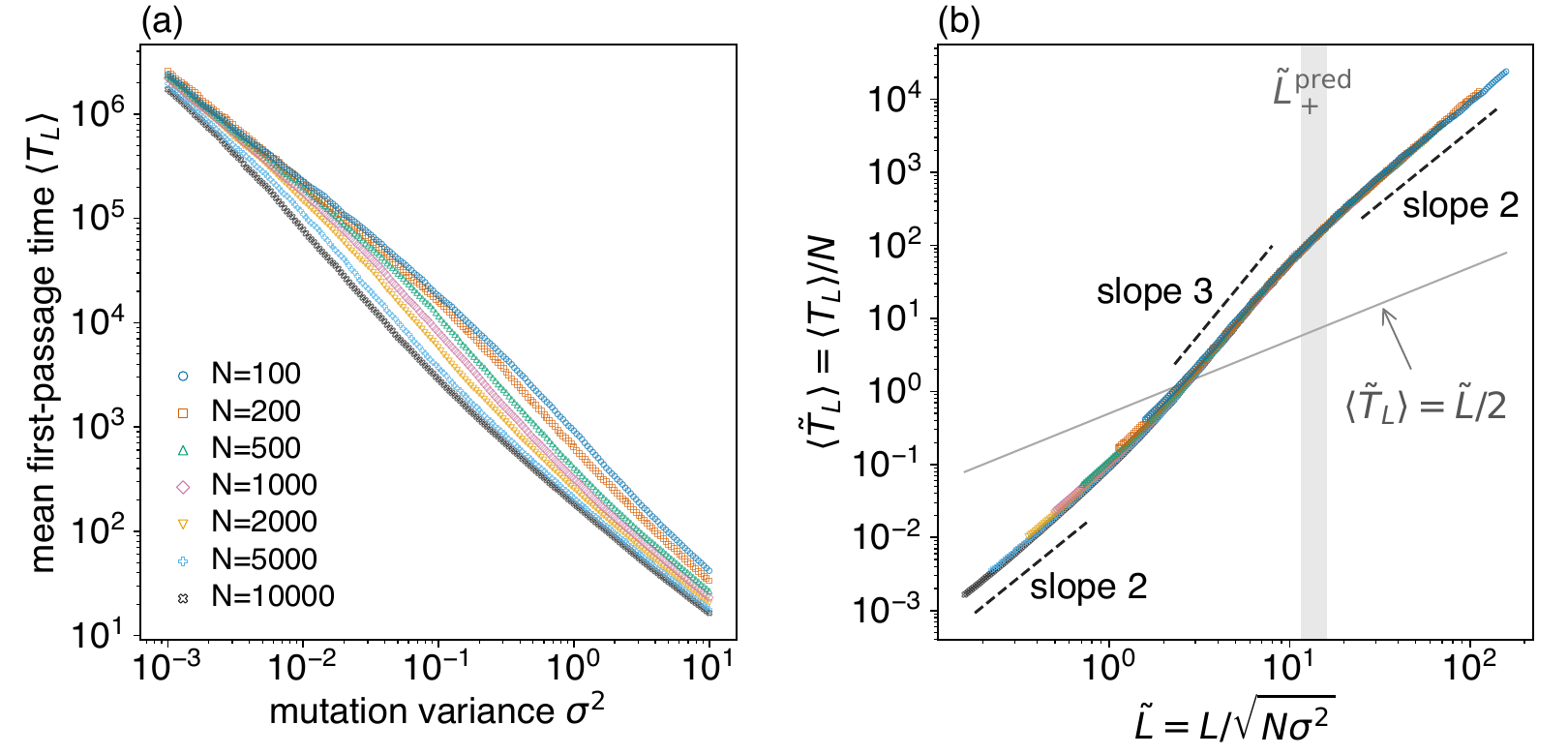}
    \caption{Scaling collapse of the first-passage law.
    (a) Mean first-passage time versus mutation variance for seven population sizes, all
    at $L=50$. (b) The same data with time rescaled by $N$ and phenotypic
    distance by $\sigma\sqrt N$.
    Open symbols represent means over 1000 realizations at each parameter
    point. Dashed segments are slope guides. The gray solid line in (b),
    $\langle\tilde T_L\rangle=\tilde L/2$, has its coefficient fixed by the leading
    lower-crossover condition [Eq.~\eqref{eq:lower_crossover}]; its
    intersections with the measured curves estimate $\tilde L_-$.
    The shaded band spans the per-population 95\% bootstrap intervals for
    $\tilde L_+^{\mathrm{pred}}=2\tilde L_-^2$
    [Eq.~\eqref{eq:crossover_relation}] across all seven population sizes,
    giving $11.5\lesssim\tilde L\lesssim16.2$. Its width reflects population-size
    variation and sampling uncertainty in the predicted position.}
    \label{fig:scaling_collapse}
\end{figure*}

\section{An anomalous first-passage law}
\label{sec:discovery}

Although each individual's phenotype follows an ordinary random walk
along its ancestral lineage, neutral evolution exhibits two crossovers
in the mean first-passage time,
with an anomalous scaling close to $(\sigma^2)^{-3/2}$ at intermediate
mutation variances and the inverse-variance law expected from diffusion
on either side [Fig.~\ref{fig:discovery_sequence}(a)].
Within the intermediate regime, reducing the mutation variance lengthens
the mean first-passage time more strongly than the inverse-variance law would
predict.
This stronger dependence is clearer in the compensated plot:
multiplying the mean by $\sigma^2/L^2$ makes an inverse-variance law
horizontal, exposing the intermediate dependence as a descending segment
with slope close to $-1/2$ [Fig.~\ref{fig:discovery_sequence}(b)].

The same pattern occurs across population sizes from $N=100$ to $10000$
[Fig.~\ref{fig:scaling_collapse}(a)]. The mutation variances at which the
crossovers occur shift with population size. To compare these data, we
introduce the rescaled distance and time
\begin{equation}
    \tilde L=\frac{L}{\sigma\sqrt N},
    \qquad \tilde t=\frac{t}{N}.
    \label{eq:fpt_scaling_form}
\end{equation}
Throughout, tildes denote dimensionless quantities.
We use the same rescaling for phenotypes, $\tilde x=x/(\sigma\sqrt N)$,
and first-passage times, $\tilde T_L=T_L/N$.
The stochastic density equation below will explain the origin of these scales.
In these units, the scaled mean first-passage times
$\langle\tilde T_L\rangle=\langle T_L\rangle/N$ for all seven population
sizes closely collapse onto a master curve when plotted against $\tilde L$
[Fig.~\ref{fig:scaling_collapse}(b)].
This curve follows a quadratic--cubic--quadratic sequence,
$\langle\tilde T_L\rangle\sim\tilde L^2\to\tilde L^3\to\tilde L^2$.
The collapse shows that the cubic law occurs over a similar intermediate
range of $\tilde L$ across population sizes, rather than being confined to
a particular population size or mutation variance.

% Float placement is handled by REVTeX in this submission layout.

\section{From reproducing individuals to collective first passage}
\label{sec:theory}

We now seek to explain the anomalous law and its two crossovers from the
reproduction and mutation rules. We first derive an equation for the
evolving distribution of phenotypes and identify the scales underlying
the collapse, then use shared ancestry to calculate the statistics of
phenotypic separation, and finally connect these results to first passage
to a target.

\subsection{From microscopic sampling to a fluctuating density}

The empirical density $\rho(x,t)=N^{-1}\sum_i\delta[x-X_i(t)]$ turns a
population realization into a field of unit total mass. Conditional on
the parental population, offspring are independent samples from its
mutation-broadened density. The one-generation covariance follows
directly from this sampling rule. For a smooth density on scales larger
than a mutational step, the resulting diffusion approximation is
\begin{equation}
    \partial_t\rho=\frac{\sigma^2}{2}\partial_x^2\rho
    +\frac{1}{\sqrt N}\eta,
    \label{eq:density_spde}
\end{equation}
with zero-mean Gaussian noise specified, conditional on the population at
time $t$, by
\begin{equation}
    \begin{aligned}
    &\left\langle\eta(x,t)\eta(y,t')\right\rangle_t\\
    &\quad=\left[\rho(x,t)\delta(x-y)-\rho(x,t)\rho(y,t)\right]
      \delta(t-t').
    \end{aligned}
    \label{eq:density_noise_covariance}
\end{equation}
This microscopic construction closely parallels the Dean--Kawasaki
description of Brownian particles \cite{Dean1996,Kawasaki1994}, and the
density equations share the same diffusion term. The noise differs
because Brownian motion obeys a local conservation law: particle numbers
change only through flux across a region's boundary. In contrast,
Wright--Fisher reproduction generates local gains and losses through fluctuations in
offspring number, so its noise is not locally conserved.
These fluctuations give the local term in
Eq.~\eqref{eq:density_noise_covariance}. The total population size
nevertheless remains fixed; the negative covariance term enforces this
global constraint by compensating an excess in one region with a deficit
elsewhere (see Supplementary Material, Sec.~S1).
Our sampling derivation recovers the diffusive-mutation case of a
general density SPDE previously derived for neutral
populations~\cite{KonnoShiga1988}.

The scales in Eq.~\eqref{eq:fpt_scaling_form} follow from these density
dynamics. Measuring time in units of $N$ and phenotypic distance in units
of $\sigma\sqrt N$, with density normalized to unit mass and the noise
rescaled accordingly, removes the explicit $N$ and $\sigma$ dependence
from both the equation and its covariance (see Supplementary Material, Sec.~S2).

At fixed spatial resolution, Eq.~\eqref{eq:density_noise_covariance}
gives a local noise variance with a term linear in density and a quadratic
normalization correction. In the dilute tail we neglect this higher-order
term, leaving local noise $\sqrt{\rho/N}\,\zeta$, with $\zeta$ standard
Gaussian white noise in phenotype space and time.

\subsection{From genealogical large deviations to an exponential tail}

First passage depends on the shape of the dilute tail as well as its local
noise. We use shared ancestry to calculate how far apart two individuals
can be in phenotype space. We then use the tail of this separation
distribution as an approximate reference for the population tail.

Let $C$ be the time to the most recent common ancestor of two distinct sampled
individuals, and $S=X_j-X_i$ their signed phenotypic separation.
The phenotype at the common ancestor cancels in this difference, so only
mutations accumulated along the two descendant branches contribute.
For a given $C=c$, each descendant branch accumulates mutational variance
$\sigma^2c$. The two branches are independent, so their variances add:
the conditional separation density $p(S\mid c)$ is Gaussian with variance
$2\sigma^2c$.
The branch duration is itself random. Two distinct lineages choose the
same parent with probability $1/N$ in each backward generation, giving
a geometric coalescence time with mean $N$ \cite{Kingman1982}.
For large $N$, this law is
approximated by the exponential density $p_C(c)=e^{-c/N}/N$.
In the long-time limit, averaging the conditional Gaussian over these
ancestral durations gives
\begin{equation}
    \begin{aligned}
    p_{\mathrm{pair}}(S)
    &\simeq\int_0^\infty dc\,
    \frac{e^{-c/N}}{N}\,
    \frac{e^{-S^2/(4\sigma^2c)}}{\sqrt{4\pi\sigma^2c}}\\
    &=\frac{1}{2\sigma\sqrt N}
      \exp\left(-\frac{|S|}{\sigma\sqrt N}\right).
    \end{aligned}
    \label{eq:pairwise_laplace}
\end{equation}
An exponential separation tail therefore emerges from Gaussian mutations
through the distribution of shared ancestral histories.
An analogous geometric separation law is known for neutral stepwise
mutations~\cite{OhtaKimura1973}.
The mean coalescence time $N$ and the exponential decay length
$\sigma\sqrt N$ recover the same time and length scales obtained from
the density dynamics.

The same genealogical calculation also determines when this tail can
form. Writing $\tilde c=c/N$ and $\tilde S=S/(\sigma\sqrt N)$, the
two exponential factors in Eq.~\eqref{eq:pairwise_laplace} express two
large-deviation costs: $\tilde c$ penalizes maintaining two distinct lineages for a long
time, while $\tilde S^2/(4\tilde c)$ penalizes producing a large
mutational separation in that time. The saddle point of this exponential
cost, $\tilde c_*=|\tilde S|/2$, selects the dominant ancestral duration.
A population initialized at a single
phenotype has only a finite time $\tilde t$ to realize that history.
For large separation and observation time at fixed ratio, the leading
large-deviation rate function is therefore obtained by constrained minimization,
\begin{equation}
    \begin{aligned}
    I_{\tilde t}(\tilde S)
    &=\min_{0<\tilde c\leq\tilde t}
      \left(\tilde c+\frac{\tilde S^2}{4\tilde c}\right)\\
    &=\begin{cases}
        |\tilde S|, & |\tilde S|\leq2\tilde t,\\[3pt]
        \displaystyle\tilde t+\frac{\tilde S^2}{4\tilde t},
          & |\tilde S|>2\tilde t.
      \end{cases}
    \end{aligned}
    \label{eq:finite_time_rate}
\end{equation}
When the optimal duration fits within the available time, the minimum
lies within the allowed interval and gives the linear cost of an exponential
tail. Otherwise the minimum lies at the endpoint $\tilde c=\tilde t$,
giving a Gaussian cost in separation. The boundary
$|\tilde S|=2\tilde t$ marks a change in the dominant ancestral history
(see Supplementary Material, Sec.~S3).

\subsection{From an exponential tail to a cubic timescale}

We now combine the exponential reference tail with the dilute-edge noise
to connect population fluctuations to first passage.

We apply the Cole--Hopf transformation $h=\log\rho$ to the local density
dynamics in the dilute tail. For a positive coarse-grained density, this
gives the formal equation
\begin{equation}
    \partial_t h
    =\frac{\sigma^2}{2}\partial_x^2h
    +\frac{\sigma^2}{2}(\partial_xh)^2
    +\frac{e^{-h/2}}{\sqrt N}\zeta+b_{\mathrm{reg}},
    \label{eq:log_density_spde}
\end{equation}
where $b_{\mathrm{reg}}$ is the It\^o correction arising from the logarithmic
transformation at fixed spatial regularization (see Supplementary Material, Sec.~S4).
The noise amplitude $e^{-h/2}/\sqrt N=1/\sqrt{N\rho}$ formally diverges
as $\rho\to0$.

In a finite population, however, an occupied bin of width $\Delta x$
contains at least one individual, so its density cannot fall below
$\rho_c=1/(N\Delta x)$. We therefore use $\rho_c$ as a one-individual
cutoff \cite{BrunetDerrida1997}, with log-density $h_c=\log\rho_c$.
At fixed coarse graining, we approximate the noise amplitude and the
It\^o correction by their values near this cutoff, while retaining
$h$ as a fluctuating field. The resulting noise amplitude
$1/\sqrt{N\rho_c}=\sqrt{\Delta x}$ is finite and independent
of $N$, and the drift becomes a constant $v_{\mathrm e}$.
Defining $\hat h(x,t)=h(x,t)-v_{\mathrm e}t$ removes this uniform drift
and gives the standard KPZ form~\cite{KPZ1986},
\begin{equation}
    \partial_t\hat h
    \simeq\frac{\sigma^2}{2}\partial_x^2\hat h
    +\frac{\sigma^2}{2}(\partial_x\hat h)^2
    +\sqrt{\Delta x}\,\zeta.
    \label{eq:edge_kpz_effective}
\end{equation}

\begin{figure}[!t]
    \centering
    \includegraphics[width=1.0\linewidth]{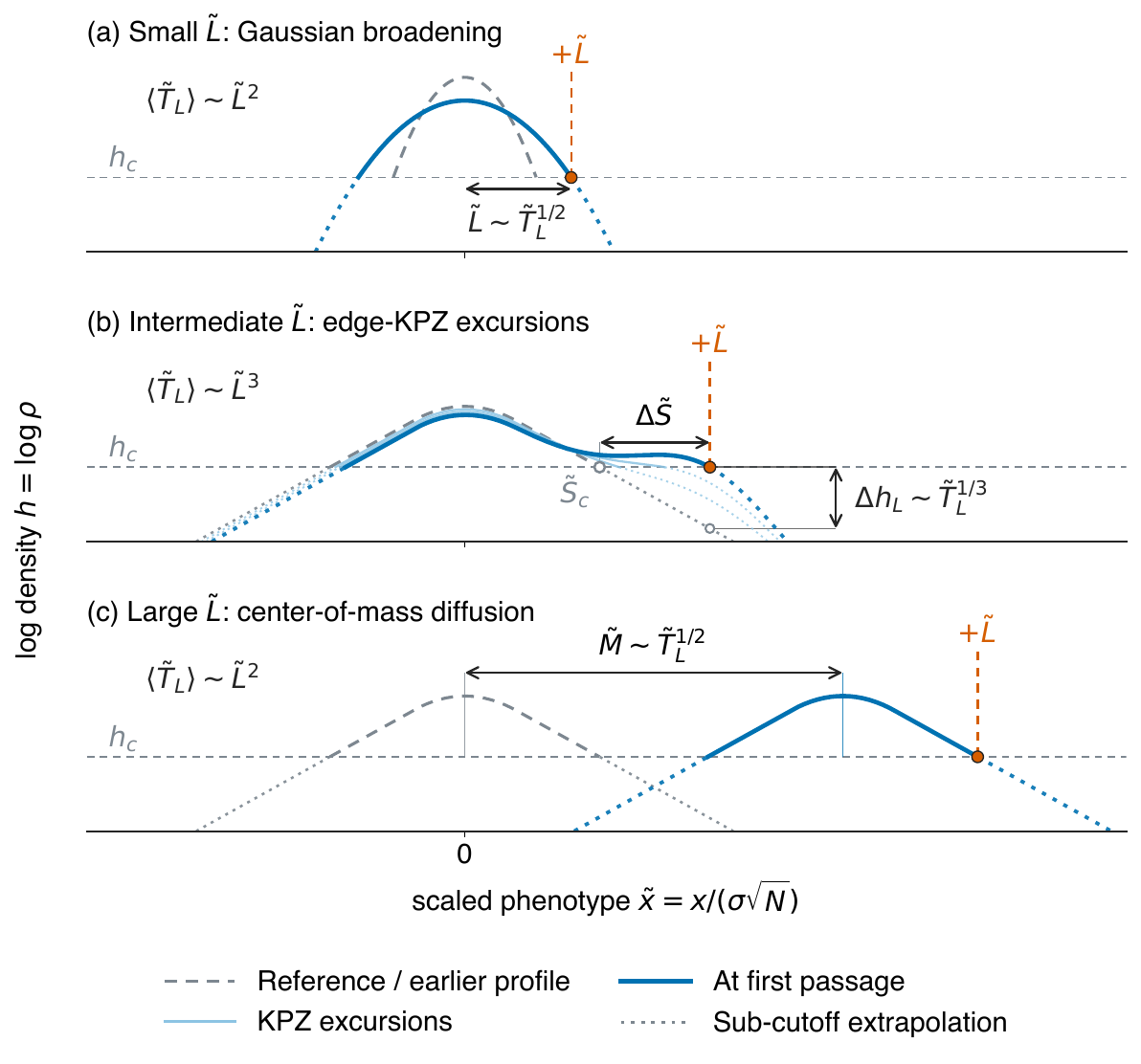}
    \caption{Mechanism for the quadratic--cubic--quadratic law.
    (a) Gaussian broadening gives quadratic scaling at small distances.
    (b) On the exponential reference tail inferred from coalescence, the
    height gap $\Delta h_L$ from the reference profile at the target to
    $h_c$ is proportional to $\Delta\tilde S=\tilde S_L-\tilde S_c$.
    In the edge-KPZ theory, height excursions growing as
    $\tilde t^{1/3}$ give a cubic passage timescale.
    (c) Center-of-mass diffusion of a cloud with finite characteristic
    width restores quadratic scaling at large distances.
    The log-density profiles share a common horizontal scale.
    Horizontal dashed lines mark the one-individual cutoff $h_c$, orange
    vertical lines mark the target, and dotted continuations are
    extrapolations below the cutoff.}
    \label{fig:three_regime_schematic}
\end{figure}

We now estimate how much the log-density at the target must rise to reach
the one-individual cutoff. We assume a smooth exponential reference tail
with the decay length $\sigma\sqrt N$ obtained in
Eq.~\eqref{eq:pairwise_laplace}, and extrapolate it from the one-individual
cutoff to the right target.
Let $\tilde S_L$ denote the scaled distance from the mean phenotype of
the background population (the bulk behind the pioneering edge) to the
target, and $\tilde S_c$ the scaled distance from the same mean
to the point where the reference profile reaches the one-individual cutoff.
Both points lie on the same exponential tail, so the common density amplitude
cancels in their log-density difference. Since the log-density has slope
$-1$ in scaled distance, the required height gap is simply the distance
between the two points [Fig.~\ref{fig:three_regime_schematic}(b)].
The uniform shift from $h$ to $\hat h$ moves the reference and cutoff
heights by the same amount, leaving this gap unchanged.

Exact solutions of the one-dimensional KPZ equation establish a
$t^{1/3}$ scale for height fluctuations at long
times~\cite{SasamotoSpohn2010,AmirCorwinQuastel2011}.
We use this growth law as a scaling hypothesis for positive height
excursions accumulated relative to this reference profile up to time
$\tilde t$.
Equating the required height gap to such an excursion at first passage gives
\begin{equation}
    \Delta h_L\simeq\tilde S_L-\tilde S_c,
    \qquad
    \Delta h_L(\tilde T_L)\sim\tilde T_L^{1/3}.
    \label{eq:height_gap_first_passage}
\end{equation}
Neglecting background displacement at leading order, the linear height
gap and the $\tilde t^{1/3}$ growth law give a cubic passage timescale.
Taking this as the characteristic timescale of the intermediate FPT
distribution, we obtain for its mean
\begin{equation}
    \langle\tilde T_L\rangle\simeq A\tilde L^3,
    \qquad
    \langle T_L\rangle\simeq A\frac{L^3}{\sigma^3\sqrt N}.
    \label{eq:cubic_dimensional}
\end{equation}
Here $A$ is a dimensionless coefficient that depends on the size of the
height excursions. Consistent with the master-curve collapse, we take
$A$ to be approximately independent of $N$ and $\sigma$ in the
intermediate regime. As the reference profile evolves,
the position where it reaches the one-individual cutoff may shift.
We allow this displacement to grow at most as $\tilde t^{1/3}$.
If the cutoff displacement and the excursion both grow as
$\tilde t^{1/3}$, their sum has the same time dependence: the cubic
distance law is preserved, with a different coefficient $A$.
A more slowly growing cutoff displacement gives a smaller correction to
that law.

\subsection{The origins of the two crossovers}

The theory also explains the physical origins of the two
crossovers. The lower crossover at scaled distance $\tilde L_-$ is set
by the time needed to form the exponential tail at the target.
The upper crossover at $\tilde L_+$ arises when diffusion of the whole
population reaches the target sooner than an edge excursion.

At separation $\tilde S$, the dominant ancestral duration in the
exponential tail is $|\tilde S|/2$.
If first passage occurs before this much time has elapsed, the relevant
tail is still Gaussian, and mutational spreading gives the short-time
quadratic regime [Fig.~\ref{fig:three_regime_schematic}(a)].
Applying the boundary $|\tilde S|=2\tilde t$ in
Eq.~\eqref{eq:finite_time_rate} at first passage determines $\tilde L_-$
through the leading condition
\begin{equation}
    2\langle\tilde T_L\rangle=\tilde L.
    \label{eq:lower_crossover}
\end{equation}

At long times, the population's characteristic width remains finite
because reproduction continually removes lineages. Two individuals in
the current population typically share an ancestor of order $N$
generations in the past, so the mutations separating them accumulate
over this timescale and give a separation of order $\sigma\sqrt N$.
The scaled mean phenotype $\tilde M$ continues to diffuse, carrying the cloud through
phenotype space~\cite{LawsonJensen2007}
(Fig.~\ref{fig:three_regime_schematic}(c); see Supplementary Material, Sec.~S5).
For targets far beyond the cloud's width, the center-of-mass approximation
gives the mean exit time
\begin{equation}
    \langle\tilde T_L\rangle_{\mathrm{CM}}\simeq\tilde L^2.
    \label{eq:center_fpt}
\end{equation}
The two quadratic regimes thus correspond to different processes:
initial broadening of the population and later diffusion of the cloud
as a whole.

Because $A\tilde L^3$ grows faster with distance than $\tilde L^2$,
center-of-mass diffusion eventually becomes the faster route to the
target. Equating the two passage times,
$A\tilde L_+^3\simeq\tilde L_+^2$, estimates the upper crossover as
$\tilde L_+\simeq1/A$.
The coefficient $A$ can in turn be estimated from the lower crossover by
substituting $\langle\tilde T_L\rangle\simeq A\tilde L^3$ into
Eq.~\eqref{eq:lower_crossover} at $\tilde L=\tilde L_-$.
These two matching steps give
\begin{equation}
    A_{\mathrm{pred}}=\frac{1}{2\tilde L_-^2},
    \qquad
    \tilde L_+^{\mathrm{pred}}=\frac{1}{A_{\mathrm{pred}}}
    =2\tilde L_-^2.
    \label{eq:crossover_relation}
\end{equation}

The cubic coefficient predicted from the lower crossover agrees with the
measured value to within about 10\% for $N\geq1000$
(see Supplementary Fig.~S1). The upper-crossover estimate, shown as the
gray band in Fig.~\ref{fig:scaling_collapse}(b), also captures the
approximate location of the broad return to quadratic behavior
(see Supplementary Material, Sec.~S6).

% Float placement is handled by REVTeX in this submission layout.

\section{Persistence under weak selection}

The anomalous first-passage law persists when the reproductive fitness
in the Wright--Fisher process depends weakly on phenotype.
We consider stabilizing selection, which favors phenotypes near the
initial value, and disruptive selection, which favors more extreme
phenotypes. The corresponding fitness weights for parental sampling are
\begin{equation}
    w_{\mathrm{stab}}(x)=e^{-\kappa x^2/2},
    \qquad
    w_{\mathrm{disr}}(x)=e^{+\kappa x^2/2},
    \label{eq:weak_selection_weights}
\end{equation}
where $\kappa=\lambda/(NL^2)$ and the dimensionless parameter $\lambda$
controls the strength of selection.
Selection is weak here because individuals near the target and those at
the origin have nearly equal expected numbers of offspring. Their relative
difference per generation is only of order $1/N$ for $\lambda$ of order unity.

\begin{figure}[tb]
    \centering
    \includegraphics[width=1.0\linewidth]{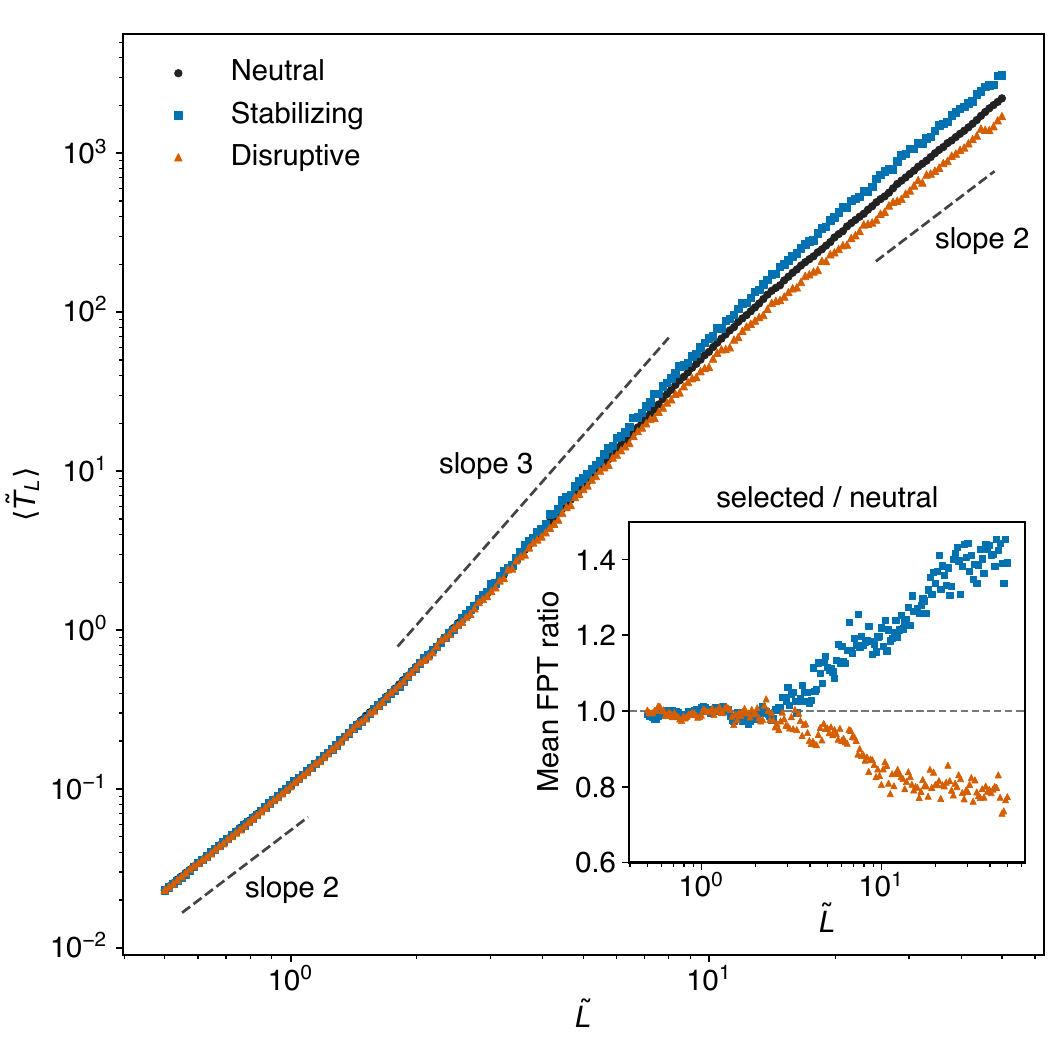}
    \caption{First-passage times under weak selection.
    Scaled mean first-passage times under neutral, weak stabilizing,
    and weak disruptive selection at $N=1000$, $L=50$, and $\lambda=1$.
    Symbols represent means over 1000 realizations per condition and
    parameter point. The inset shows each selected mean divided by the
    neutral mean at the same mutation variance; its horizontal line marks
    a ratio of one. Dashed segments in the main plot are guides for slopes
    2, 3, and 2.}
    \label{fig:weak_selection}
\end{figure}

Weak selection leaves the mean first-passage time almost unchanged in the
short-time regime of Gaussian spreading [Fig.~\ref{fig:weak_selection}].
As the scaled distance increases through the anomalous regime, deviations
from the neutral mean gradually become more pronounced: stabilizing
selection generally lengthens the waiting time, while disruptive selection
generally shortens it. The deviations are largest in the long-time regime
dominated by motion of the population as a whole, as seen in the inset.
Despite these shifts, both forms of selection preserve the three-regime
structure, with an intermediate law close to
$\langle T_L\rangle\propto(\sigma^2)^{-3/2}$ between two approximately
inverse-variance regimes.

\section{Discussion}
\label{sec:discussion}

In this paper, we have shown that Wright--Fisher evolution exhibits an
anomalous collective first-passage law under neutral and weakly selected
conditions, even though the mutational steps are diffusive.
The mean waiting time scales approximately as $(\sigma^2)^{-3/2}$ in
an intermediate regime bounded by two crossovers to the usual
inverse-variance dependence.
To explain this behavior, we developed an edge-KPZ theory combining
genealogical statistics with a microscopic fluctuating-density
description. By connecting the exponential reference tail to fluctuations
at the population edge, this theory explains the cubic dependence on
target distance and relates its amplitude to the two crossover scales.

In our theory, reproduction generates correlations among individuals
through shared ancestry, giving rise to an edge-KPZ mechanism for
anomalous first passage.
This raises the possibility that, in correlated diffusive systems beyond
evolution, KPZ fluctuations in dilute tails provide a common route to
nontrivial first-passage statistics.
Indeed, KPZ fluctuations in first-arrival times have been predicted for
particles whose trajectories are correlated by a shared time-dependent
random environment \cite{LeDoussalThiery2017}.
In a large-distance regime of such shared-environment models, the
environment produces anomalous arrival-time fluctuations
\cite{HassDrillickCorwinCorwin2024}, but the mean first-passage time does
not exhibit the anomalous scaling found here.
A theory of how different sources of correlation affect edge-KPZ
dynamics could clarify when the mean first-passage law becomes anomalous
and how the distribution of arrival times changes.
Such a study would extend edge-KPZ theory to explain anomalous diffusion
across a broad range of systems beyond evolution.

Our theory could help explain how rapidly populations gain access to
new phenotypes through neutral exploration of genotypes that preserve
the existing phenotype \cite{Wagner2008}.
Previous work relates population distributions to network topology
\cite{vanNimwegen1999} and estimates phenotype discovery times by
replacing local mutational neighborhoods with a network average
\cite{SchaperLouis2014}.
Nearby genotypes can, however, give access to much the same set of
phenotypes through mutation.
A population exploring such a region may therefore need to reach a
different part of the network before a target phenotype becomes
accessible.
Averaging over the network can miss the time required for this search.
A quantitative description of this exploration remains incomplete for
populations of intermediate genetic diversity \cite{Martin2024}.
Earlier simulations of neutral populations in finite, discrete genotype
spaces found an anomalous dependence of the mean discovery time on mutation
rate~\cite{vanNimwegenCrutchfield2000}.
These findings motivate extending edge-KPZ theory to genotype networks
by connecting genealogical timescales and fluctuations in sparsely
occupied regions to the rate of exploration.
A central challenge is to identify the counterpart of a dilute
population edge on such networks and determine whether its fluctuations
admit an effective KPZ description.
Developing this connection could clarify how shared ancestry and
population fluctuations shape the timescale of evolutionary innovation
and its dependence on population size and mutation rate.

Predictions from the edge-KPZ theory can be readily tested experimentally.
Measurements should reveal an anomalous first-passage regime bounded
by two crossovers to ordinary diffusive scaling.
Although mutator strains spanning a broad range of mutation rates are
available, varying the mutation rate without changing the mutation
spectrum or growth rate can be experimentally challenging
\cite{Shibai2025}.
The simplest test therefore varies only the target distance, keeping
population size and mutational conditions fixed.
Microbial serial-transfer experiments already provide controlled
population-size conditions \cite{Desai2007,Rozen2008}.
Phenotypes could be measured at regular generation intervals, recording
the individual furthest from the initial phenotype.
With sufficiently frequent measurements and sensitivity to rare
individuals, the same record would allow first-passage times for many
target distances to be estimated retrospectively.
Independent replicate populations would then provide the mean
first-passage curve without separate evolution experiments for each
target.
The persistence of this law under weak selection broadens the range of
experimental systems in which it can be tested, without requiring an
exactly neutral fitness landscape.

Revisiting a fundamental question in a classical model of neutral
evolution has uncovered an anomalous law for the appearance of new
phenotypes.
The edge-KPZ theory connects this basic evolutionary problem to the
statistical physics of fluctuating growth through shared ancestry and
the dilute population edge.
This connection lays the foundation for a broader theory of the pace
of evolution, opening new possibilities for predicting the emergence
of new phenotypes and the timescales of evolutionary innovation.

% \section*{Acknowledgments}

% TODO: Add funding, computational resources, and acknowledgments.

% TODO: Add the repository URL and the archived-data location before submission.

% TODO: Restore the bibliography after references have been added to refs.bib.
% \bibliographystyle{unsrt}
% \bibliography{refs}
% The bibliography is enabled below for the reconstructed manuscript.
\bibliography{refs}

\clearpage
\onecolumngrid
\setcounter{section}{0}
\setcounter{subsection}{0}
\setcounter{equation}{0}
\setcounter{figure}{0}
\setcounter{table}{0}
\renewcommand{\thesection}{S\arabic{section}}
\renewcommand{\thesubsection}{S\arabic{section}.\arabic{subsection}}
\renewcommand{\theequation}{S\arabic{equation}}
\renewcommand{\thefigure}{S\arabic{figure}}
\renewcommand{\thetable}{S\arabic{table}}
\renewcommand{\theHsection}{S\arabic{section}}
\renewcommand{\theHsubsection}{S\arabic{section}.\arabic{subsection}}
\renewcommand{\theHequation}{S\arabic{equation}}
\renewcommand{\theHfigure}{S\arabic{figure}}
\renewcommand{\theHtable}{S\arabic{table}}
\makeatletter
\renewcommand{\p@subsection}{}
\makeatother

\section*{Supplementary Material}

\section{Density fluctuations generated by Wright--Fisher sampling}
\label{sec:supp_spde}

\subsection{One-generation sampling covariance}

Consider a neutral population with phenotypes $X_1(t),\ldots,X_N(t)$,
starting from the monomorphic state $X_i(0)=0$.  Define the empirical
density
\begin{equation}
    \rho(x,t)=\frac{1}{N}\sum_{i=1}^{N}\delta[x-X_i(t)],
    \qquad
    \int \rho(x,t)\,dx=1.
    \label{eq:supp_empirical_density}
\end{equation}
We use angle brackets for ensemble averages. The notation
$\langle\cdot\rangle_t$ denotes an average over the parental choices and
mutations that generate the next generation, with the entire current
population $\{X_i(t)\}$ held fixed. Brackets without the subscript denote
averages over population histories.
Each offspring chooses a parent uniformly from the preceding generation and
receives a Gaussian mutation with mean zero and variance $\sigma^2$.
Its exact conditional density is the Gaussian convolution
\[
    P(x,t)=\frac{1}{N}\sum_{i=1}^N
    \frac{1}{\sqrt{2\pi\sigma^2}}
    \exp\left[-\frac{(x-X_i(t))^2}{2\sigma^2}\right].
\]
Conditional on the current population, $P$ is fixed and the $N$ offspring
are independent draws from it. The realized next-generation density
fluctuates around this sampling distribution.

To calculate these fluctuations, let $\phi(x)$ be an arbitrary smooth
test function and define the weighted population average
\begin{equation}
    I_\phi=\int \phi(x)\rho(x,t+1)\,dx
    =\frac{1}{N}\sum_{i=1}^{N}\phi[X_i(t+1)].
    \label{eq:supp_test_function}
\end{equation}
For example, $\phi(x)=x$ gives the mean phenotype of the next generation,
while a smooth window function measures a weighted population fraction
in a chosen phenotype interval. Test functions therefore let us examine
the fluctuating density through ordinary population observables, without
evaluating its delta peaks pointwise. Since $I_\phi$ is a sample average,
its conditional mean is
\begin{equation}
    \langle I_\phi\rangle_t
    =\frac{1}{N}\sum_{i=1}^N\int\phi(x)P(x,t)\,dx
    =\int\phi(x)P(x,t)\,dx.
    \label{eq:supp_test_mean}
\end{equation}
To determine the covariance, introduce a second test function $\psi$.
In the double sum for $\langle I_\phi I_\psi\rangle_t$, the $N$ terms
with $i=j$ evaluate both functions on the same offspring. The
$N(N-1)$ terms with $i\ne j$ factor into products of means because
different offspring are conditionally independent. Thus,
\[
    \begin{aligned}
    \langle I_\phi I_\psi\rangle_t
    &=\frac{1}{N^2}\sum_{i,j=1}^N
      \big\langle\phi[X_i(t+1)]\psi[X_j(t+1)]\big\rangle_t\\
    &=\frac{1}{N^2}\sum_{i=1}^N
      \big\langle\phi[X_i(t+1)]\psi[X_i(t+1)]\big\rangle_t\\
    &\quad+\frac{1}{N^2}\sum_{i\ne j}
      \big\langle\phi[X_i(t+1)]\big\rangle_t
      \big\langle\psi[X_j(t+1)]\big\rangle_t\\
    &=\frac{1}{N}\int\phi(x)\psi(x)P(x,t)\,dx
      +\left(1-\frac{1}{N}\right)
       \langle I_\phi\rangle_t\langle I_\psi\rangle_t.
    \end{aligned}
\]
Writing $\delta I_\phi=I_\phi-\langle I_\phi\rangle_t$ and subtracting
the product of means gives
\[
    \langle\delta I_\phi\delta I_\psi\rangle_t
    =\frac{1}{N}\left\{
       \int\phi(x)\psi(x)P(x,t)\,dx
       -\left[\int\phi(x)P(x,t)\,dx\right]
        \left[\int\psi(y)P(y,t)\,dy\right]
      \right\}.
\]
The factor $1/N$ results from averaging $N$ independent offspring: after
subtracting the product of means, only the same-offspring covariances
remain in the double sum.
In particular, $\psi=\phi$ gives the conditional variance,
\begin{equation}
    \langle(\delta I_\phi)^2\rangle_t
    =\frac{1}{N}\left\{
        \int \phi(x)^2P(x,t)\,dx
        -\left[\int \phi(x)P(x,t)\,dx\right]^2
    \right\}.
    \label{eq:supp_test_variance}
\end{equation}

We now recover the density fluctuations from these weighted averages.
Define $\xi(x,t)=\rho(x,t+1)-P(x,t)$. Its conditional mean vanishes,
$\langle\xi(x,t)\rangle_t=0$, because $P$ is the mean offspring density
at fixed current population. Subtracting the mean of the weighted
observable therefore isolates the same fluctuation:
\[
    \begin{aligned}
    \delta I_\phi
    &=\int\phi(x)\rho(x,t+1)\,dx-\int\phi(x)P(x,t)\,dx\\
    &=\int\phi(x)\xi(x,t)\,dx.
    \end{aligned}
\]
Taking the product for two test functions and averaging gives
\[
    \begin{aligned}
    \langle\delta I_\phi\delta I_\psi\rangle_t
    &=\left\langle
      \left[\int\phi(x)\xi(x,t)\,dx\right]
      \left[\int\psi(y)\xi(y,t)\,dy\right]
      \right\rangle_t\\
    &=\iint\phi(x)\psi(y)
      \langle\xi(x,t)\xi(y,t)\rangle_t\,dx\,dy.
    \end{aligned}
\]
To compare this expression with the covariance obtained from the
individual samples, we write both terms in that result as double
integrals:
\[
    \begin{aligned}
    \int\phi(x)\psi(x)P(x,t)\,dx
    &=\iint\phi(x)\psi(y)P(x,t)\delta(x-y)\,dx\,dy,\\
    \left[\int\phi(x)P(x,t)\,dx\right]
    \left[\int\psi(y)P(y,t)\,dy\right]
    &=\iint\phi(x)\psi(y)P(x,t)P(y,t)\,dx\,dy.
    \end{aligned}
\]
The delta function in the first identity enforces that both functions
are evaluated at the same phenotype. The product of means instead
involves two separately integrated phenotypes. The covariance of these population averages
therefore has the same weighted double-integral form as the noise
covariance, with kernel $[P(x,t)\delta(x-y)-P(x,t)P(y,t)]/N$.
Since $\phi$ and $\psi$ can be chosen independently and arbitrarily,
agreement for all such weighted measurements fixes the spatial
covariance, in the distributional sense:
\begin{equation}
    \langle\xi(x,t)\xi(y,t)\rangle_t
    =\frac{1}{N}\left[
        P(x,t)\delta(x-y)-P(x,t)P(y,t)
    \right].
    \label{eq:supp_sampling_covariance}
\end{equation}
Taking $\phi=1$ gives $I_1=1$ in every realization and therefore
$\delta I_1=0$. Since $\delta I_1=\int\xi(x,t)\,dx$, the total-mass
fluctuation vanishes. Integrating the covariance over either coordinate
also gives zero: $\int P(x,t)\,dx=1$ makes the local and nonlocal terms
cancel.
This result is exact for one discrete generation at finite
$N$ and includes the randomness of both parental choice and mutation.
No Gaussian approximation for $\xi$ has been used.

\subsection{Diffusion approximation and the dilute edge}

The exact one-generation update separates the conditional mean change
from the sampling fluctuation:
\[
    \rho(x,t+1)-\rho(x,t)=P(x,t)-\rho(x,t)+\xi(x,t).
\]
For a coarse-grained density varying smoothly over a mutational step,
expanding the Gaussian convolution gives
\begin{equation}
    P(x,t)=\rho(x,t)+\frac{\sigma^2}{2}\partial_x^2\rho(x,t)
    +\mathcal{O}(\sigma^4\partial_x^4\rho).
    \label{eq:supp_expected_density}
\end{equation}
The linear term in the mutation increment averages to zero, while its
second moment $\sigma^2$ supplies the diffusive term. In the covariance
of Eq.~\eqref{eq:supp_sampling_covariance}, we likewise retain the leading
term by replacing $P$ with $\rho$.

Fresh sampling in each generation supplies conditionally centered
increments with covariance of order $1/N$. Each increment has zero
conditional mean given its past, so sampling fluctuations from different
generations are uncorrelated. On timescales long compared with one
generation, we approximate these increments by Gaussian white
noise with the same leading conditional mean and covariance. The resulting
continuous-time description is interpreted in the It\^o sense: the noise
amplitude is evaluated at the density immediately before the increment.
With one generation as the unit of time, the formal diffusion
approximation is
\begin{equation}
    \partial_t\rho(x,t)
    =\frac{\sigma^2}{2}\partial_x^2\rho(x,t)
    +\frac{1}{\sqrt{N}}\eta(x,t),
    \label{eq:supp_density_spde_full}
\end{equation}
with zero-mean Gaussian noise whose instantaneous conditional covariance is
\begin{equation}
    \left\langle\eta(x,t)\eta(y,t')\right\rangle_t
    =\left[\rho(x,t)\delta(x-y)-\rho(x,t)\rho(y,t)\right]
    \delta(t-t').
    \label{eq:supp_density_noise_covariance}
\end{equation}
The subscript $t$ again conditions on the current population. The
$1/N$ sampling covariance gives the noise prefactor $N^{-1/2}$.

At fixed spatial resolution, the local noise variance has a term linear
in density and a quadratic normalization correction. In the dilute tail,
the latter is subleading and can be neglected. Introducing standard
spatiotemporal Gaussian white noise $\zeta$ gives the local approximation
\begin{equation}
    \partial_t\rho
    \simeq\frac{\sigma^2}{2}\partial_x^2\rho
    +\sqrt{\frac{\rho}{N}}\,\zeta,
    \qquad
    \langle\zeta(x,t)\zeta(y,t')\rangle
    =\delta(x-y)\delta(t-t').
    \label{eq:supp_density_spde_edge}
\end{equation}

\section{Characteristic scales and finite-size scaling}
\label{sec:supp_nondimensionalization}

Set
\begin{equation}
    x=\alpha\tilde{x},
    \qquad
    t=\beta\tilde{t},
    \qquad
    \rho(x,t)=\frac{1}{\alpha}\tilde{\rho}(\tilde{x},\tilde{t}).
    \label{eq:supp_nondim_definition}
\end{equation}
As in the main text, tildes denote dimensionless quantities.
The density rescaling preserves unit total mass. The spatial and temporal
delta functions in the white-noise covariance contribute factors
$\alpha^{-1}$ and $\beta^{-1}$, respectively, so white noise transforms as
\begin{equation}
    \zeta(x,t)=\frac{1}{\sqrt{\alpha\beta}}
    \tilde{\zeta}(\tilde{x},\tilde{t}).
    \label{eq:supp_noise_rescaling}
\end{equation}
Substitution into Eq.~\eqref{eq:supp_density_spde_edge} yields
\begin{equation}
    \partial_{\tilde{t}}\tilde{\rho}
    =\frac{\beta\sigma^2}{2\alpha^2}
        \partial_{\tilde{x}}^2\tilde{\rho}
    +\sqrt{\frac{\beta}{N}}\sqrt{\tilde{\rho}}\,
        \tilde{\zeta}.
    \label{eq:supp_spde_rescaled}
\end{equation}
Choosing
\begin{equation}
    \beta=N,
    \qquad
    \alpha=\sigma\sqrt{N}
    \label{eq:supp_characteristic_scales}
\end{equation}
removes the parameter dependence from the coefficients of the local
continuum equation:
\begin{equation}
    \partial_{\tilde{t}}\tilde{\rho}
    =\frac{1}{2}\partial_{\tilde{x}}^2\tilde{\rho}
    +\sqrt{\tilde{\rho}}\,\tilde{\zeta}.
    \label{eq:supp_spde_dimensionless}
\end{equation}
The full covariance in Eq.~\eqref{eq:supp_density_noise_covariance} has
the same scaling: both terms in its spatial kernel acquire a factor
$\alpha^{-2}$, while $\delta(t-t')$ contributes $\beta^{-1}$.
Defining $\eta(x,t)=(\alpha\sqrt\beta)^{-1}
\tilde\eta(\tilde x,\tilde t)$ therefore preserves the covariance
structure and gives the same coefficients $1/2$ and $1$ in the full
rescaled density equation.

For a target at distance $L$, these rescalings give
\begin{equation}
    \langle\tilde{T}_L\rangle=\frac{\langle T_L\rangle}{N},
    \qquad
    \tilde{L}=\frac{L}{\sigma\sqrt{N}}.
    \label{eq:supp_fpt_scaling_form}
\end{equation}
The continuum scaling predicts a master curve of
$\langle\tilde T_L\rangle$ against $\tilde L$.
Discrete generations and finite population size can produce corrections
to this collapse.

\section{Finite-time coalescent distribution of phenotypic separation}
\label{sec:supp_coalescent}

\subsection{Long-time pairwise distribution}

Let $C$ be the backward time to the most recent common ancestor of two
distinct individuals sampled from the same generation.  At each backward
step, two distinct lineages choose the same parent with probability
$1/N$.  Coalescence at step $c$ requires $c-1$ steps without coalescence,
followed by a shared parent.  Hence
\begin{equation}
    \Pr(C=c)=\frac{1}{N}\left(1-\frac{1}{N}\right)^{c-1}.
    \label{eq:supp_coalescence_geometric}
\end{equation}
Thus, $\langle C\rangle=N$.  This is an exact discrete probability law.

Conditional on $C=c$, each descendant branch contains $c$ independent
mutational increments, so its displacement from the common ancestor has
variance $\sigma^2c$.  Taking the difference removes the common ancestral
phenotype and adds the variances of the two independent branches.
The signed separation $S=X_j-X_i$ therefore obeys
\begin{equation}
    S\mid C=c\sim\mathcal{N}(0,2\sigma^2c),
    \label{eq:supp_conditional_separation}
\end{equation}
and hence
\begin{equation}
    p(S\mid c)=\frac{1}{\sqrt{4\pi\sigma^2c}}
    \exp\left(-\frac{S^2}{4\sigma^2c}\right).
    \label{eq:supp_conditional_separation_density}
\end{equation}
In the long-time limit, the unconditional density is the exact mixture
of these Gaussian distributions over the possible coalescence times:
\[
    p_{\mathrm{pair}}(S)
    =\sum_{c=1}^{\infty}\Pr(C=c)\,p(S\mid c).
\]
Each term is the probability of an ancestral duration multiplied by the
separation density at that duration.

For large $N$, we approximate the coalescence time by a continuous
exponential waiting time with density
\begin{equation}
    p_C(c)=\frac{1}{N}e^{-c/N},\qquad c>0.
    \label{eq:supp_coalescence_exponential}
\end{equation}
This is the two-lineage limit of the neutral coalescent~\cite{Kingman1982}.
Replacing the discrete waiting-time law by $p_C(c)$ turns the sum into
an integral,
\begin{align}
    p_{\mathrm{pair}}(S)
    &\simeq\int_0^\infty
    \frac{e^{-c/N}}{N}\,
    \frac{e^{-S^2/(4\sigma^2c)}}{\sqrt{4\pi\sigma^2c}}\,dc \notag\\
    &=\frac{1}{2\sigma\sqrt{N}}e^{-|S|/(\sigma\sqrt{N})}.
    \label{eq:supp_pairwise_laplace}
\end{align}
Here we used the integral identity
\[
    \int_0^\infty c^{-1/2}e^{-ac-b/c}\,dc
    =\sqrt{\frac{\pi}{a}}\,e^{-2\sqrt{ab}},
    \qquad a>0,\quad b\geq0,
\]
with $a=1/N$ and $b=S^2/(4\sigma^2)$.

\subsection{Finite observation time}

To extend this calculation to finite observation times, we next identify
the ancestral duration that dominates the integral at large separations.
Writing $\tilde c=c/N$ and $\tilde S=S/(\sigma\sqrt N)$, the two factors
in the integrand combine as
\begin{align}
    p_{\mathrm{pair}}(S)
    &\simeq\frac{1}{\sigma\sqrt N}
    \int_0^\infty\frac{d\tilde c}{\sqrt{4\pi\tilde c}}\,
    e^{-\Phi(\tilde c;\tilde S)}, \notag\\
    \Phi(\tilde{c};\tilde{S})
    &=\tilde{c}+\frac{\tilde{S}^2}{4\tilde{c}}.
    \label{eq:supp_coalescent_action}
\end{align}
Thus, $\Phi$ is the exponential cost of a contribution at fixed ancestral
duration $\tilde c$. At large $|\tilde S|$, the dominant contribution
comes from its minimum. The stationarity condition
$\partial_{\tilde c}\Phi=1-\tilde S^2/(4\tilde c^2)=0$ gives
\begin{equation}
    \tilde{c}_* = \frac{|\tilde{S}|}{2},
    \qquad
    \Phi(\tilde{c}_*;\tilde{S})=|\tilde{S}|.
    \label{eq:supp_coalescent_saddle}
\end{equation}
To see why the integral concentrates near this minimum, write
$\delta\tilde c=\tilde c-\tilde c_*$.
The cost above its minimum is
\[
    \Phi(\tilde c_*+\delta\tilde c;\tilde S)-|\tilde S|
    =\frac{(\delta\tilde c)^2}{\tilde c_*+\delta\tilde c}
    =\frac{2(\delta\tilde c)^2}{|\tilde S|}
    +\mathcal O\!\left(\frac{|\delta\tilde c|^3}{|\tilde S|^2}\right).
\]
A fixed fractional displacement from $\tilde c_*$ therefore incurs an
additional cost proportional to $|\tilde S|$ and is exponentially
suppressed. The quadratic term gives appreciable contributions over a width of order
$|\tilde S|^{1/2}$, whose ratio to $\tilde c_*=|\tilde S|/2$ vanishes as
$|\tilde S|^{-1/2}$. Across this range, both the omitted terms in the
exponent and the relative variation of the prefactor $\tilde c^{-1/2}$
are of order $|\tilde S|^{-1/2}$. These vanishing corrections justify
using $\Phi(\tilde c_*;\tilde S)$ for the leading exponential dependence
when $|\tilde S|\gg1$.

For a population started monomorphic at $t=0$, the available mutational
duration at observation time $t=N\tilde t>0$ is $\min(C,t)$.
In the same continuous approximation, the probability that the lineages
remain distinct back to the initial population is $e^{-\tilde t}$.
Their mutation histories then both start at the common initial phenotype,
giving separation density $p(S\mid t)$.
Splitting the mixture at $C=t$ therefore gives
\begin{equation}
    p_{\mathrm{pair}}(S,t)
    \simeq\frac{1}{\sigma\sqrt{N}}
    \left[
    \int_0^{\tilde{t}}
        \frac{e^{-\Phi(\tilde c;\tilde S)}}{\sqrt{4\pi\tilde{c}}}
        \,d\tilde{c}
    +\frac{e^{-\Phi(\tilde t;\tilde S)}}{\sqrt{4\pi\tilde{t}}}
    \right].
    \label{eq:supp_pairwise_finite_continuum}
\end{equation}
The integral and the initial-time contribution carry total weights
$1-e^{-\tilde t}$ and $e^{-\tilde t}$, respectively.

The saddle in Eq.~\eqref{eq:supp_coalescent_saddle} controls the integral
when it lies within $0<\tilde c\leq\tilde t$. Otherwise, both the
integral and the initial-time contribution have the leading exponent
$\Phi(\tilde t;\tilde S)$. Within this continuum approximation, for large
separation and observation time at fixed ratio,
$-\log[\sigma\sqrt N\,p_{\mathrm{pair}}(S,t)]
\simeq I_{\tilde t}(\tilde S)$ to leading order, where
\begin{equation}
    \begin{aligned}
    I_{\tilde{t}}(\tilde{S})
    &=\min_{0<\tilde c\leq\tilde t}\Phi(\tilde c;\tilde S)\\
    &=
    \begin{cases}
        |\tilde{S}|, & |\tilde{S}|\leq2\tilde{t},\\[3pt]
        \displaystyle\tilde{t}+\frac{\tilde{S}^2}{4\tilde{t}},
        & |\tilde{S}|>2\tilde{t}.
    \end{cases}
    \end{aligned}
    \label{eq:supp_finite_time_rate}
\end{equation}
At finite $\tilde S$ and $\tilde t$, the change between the two
asymptotic branches is smooth.

\section{Log-density dynamics near the one-individual cutoff}
\label{sec:supp_pair_to_edge}
\label{sec:supp_kpz_mapping}

For a positive coarse-grained density, set $h=\log\rho$.
This Cole--Hopf transformation gives $\partial_x\rho=\rho\,\partial_xh$
and hence
\[
    \frac{\partial_x^2\rho}{\rho}
    =\partial_x^2h+(\partial_xh)^2.
\]
Applying It\^o's formula to Eq.~\eqref{eq:supp_density_spde_edge}
therefore gives the formal equation
\begin{equation}
    \partial_t h
    =\frac{\sigma^2}{2}\partial_x^2h
    +\frac{\sigma^2}{2}(\partial_xh)^2
    +\frac{e^{-h/2}}{\sqrt{N}}\zeta+b_{\mathrm{reg}}.
    \label{eq:supp_log_density_spde}
\end{equation}
Here $b_{\mathrm{reg}}$ is the It\^o correction at fixed spatial
regularization. The noise amplitude follows by dividing the density
noise by $\rho$:
\[
    \frac{\sqrt{\rho/N}}{\rho}
    =\frac{1}{\sqrt{N\rho}}
    =\frac{e^{-h/2}}{\sqrt N}.
\]

To evaluate the It\^o correction, use the local noise approximation in
Eq.~\eqref{eq:supp_density_spde_edge}.
For a bin of width $\Delta x$ with coarse-grained occupancy
$n=N\rho\Delta x>0$, the local sampling increment is
$dn|_{\mathrm{sampling}}=\sqrt n\,dW_t$, where $W_t$ is a standard
Wiener process with $\langle dW_t\rangle=0$ and
$\langle(dW_t)^2\rangle=dt$. Its contribution to the logarithmic
increment is therefore
\[
    d\log n\big|_{\mathrm{sampling}}
    =\frac{dW_t}{\sqrt n}-\frac{dt}{2n}.
\]
At fixed $N$ and $\Delta x$, $h=\log n-\log(N\Delta x)$, so
$dh=d\log n$. The term $-dt/(2n)$ therefore gives
$b_{\mathrm{reg}}=-1/(2n)=-1/(2N\rho\Delta x)$ in the local noise
approximation.

We approximate the noise amplitude and the It\^o correction by their
values at the one-individual cutoff
$\rho_c=1/(N\Delta x)$~\cite{BrunetDerrida1997},
while retaining $h(x,t)$ as a fluctuating field.
The noise amplitude becomes
\begin{equation}
    \frac{1}{\sqrt{N\rho_c}}=\sqrt{\Delta x}.
    \label{eq:supp_cutoff_noise_strength}
\end{equation}
The noise averaged over disjoint bins obeys
$\langle\zeta_j(t)\zeta_k(t')\rangle
=\delta_{jk}\delta(t-t')/\Delta x$.
Consequently, $\sqrt{\Delta x}\,\zeta_j$ has unit variance rate in
each bin, expressing the order-one relative noise at one-individual
occupancy. Evaluating the local It\^o correction at the same cutoff
gives a constant $v_{\mathrm e}=-1/2$ per generation.
Although this drift is not small, subtracting $v_{\mathrm e}t$ from both
the reference log-density and the cutoff height leaves their difference
unchanged.

With both coefficients evaluated at the cutoff, the effective equation is
\begin{equation}
    \partial_t h
    \simeq\frac{\sigma^2}{2}\partial_x^2h
    +\frac{\sigma^2}{2}(\partial_xh)^2
    +\sqrt{\Delta x}\,\zeta+v_{\mathrm e}.
    \label{eq:supp_log_density_cutoff}
\end{equation}
To write the standard KPZ form~\cite{KPZ1986} used in the main text, define
$\hat h(x,t)=h(x,t)-v_{\mathrm e}t$.  Equation~\eqref{eq:supp_log_density_cutoff}
then becomes
\[
    \partial_t\hat h
    \simeq\frac{\sigma^2}{2}\partial_x^2\hat h
    +\frac{\sigma^2}{2}(\partial_x\hat h)^2
    +\sqrt{\Delta x}\,\zeta.
\]

\section{Exact neutral moments}
\label{sec:supp_center_moments}

\subsection{Within-population variance}

Define the population center of mass (mean phenotype) and the
within-population variance by
\begin{equation}
    M_t=\frac{1}{N}\sum_{i=1}^{N}X_i(t),
    \qquad
    V_t=\frac{1}{N}\sum_{i=1}^{N}[X_i(t)-M_t]^2.
    \label{eq:supp_mean_variance}
\end{equation}
Conditional on the current generation, write the offspring phenotypes as
\begin{equation}
    Y_i=X_{A_i}(t)+\sigma Z_i,
    \qquad Z_i\sim\mathcal{N}(0,1),
    \label{eq:supp_offspring_update}
\end{equation}
where the $A_i$ are independent uniform parental choices.  Conditional
averages satisfy
\begin{equation}
    \langle Y_i\rangle_t=M_t,
    \qquad
    \langle(Y_i-M_t)^2\rangle_t=V_t+\sigma^2.
    \label{eq:supp_offspring_moments}
\end{equation}
Because offspring are conditionally independent,
\begin{equation}
    \langle(M_{t+1}-M_t)^2\rangle_t
    =\frac{V_t+\sigma^2}{N}.
    \label{eq:supp_center_increment_variance}
\end{equation}

The offspring mean-square displacement from the old center separates
into the variance around the new center and the squared shift of that center:
\begin{equation}
    \frac{1}{N}\sum_{i=1}^{N}(Y_i-M_t)^2
    =V_{t+1}+(M_{t+1}-M_t)^2.
    \label{eq:supp_variance_decomposition}
\end{equation}
Taking the conditional average and using
Eqs.~\eqref{eq:supp_offspring_moments} and
\eqref{eq:supp_center_increment_variance} gives
\begin{equation}
    \langle V_{t+1}\rangle_t
    =\left(1-\frac{1}{N}\right)(V_t+\sigma^2).
    \label{eq:supp_variance_recursion_conditional}
\end{equation}
Averaging over the current populations removes the conditioning,
$\langle\langle V_{t+1}\rangle_t\rangle=\langle V_{t+1}\rangle$, and gives
\[
    \langle V_{t+1}\rangle
    =\left(1-\frac{1}{N}\right)
     \big(\langle V_t\rangle+\sigma^2\big).
\]
Iterating from $V_0=0$ yields
\begin{equation}
    \langle V_t\rangle
    =(N-1)\sigma^2
    \left[1-\left(1-\frac{1}{N}\right)^t\right].
    \label{eq:supp_variance_solution}
\end{equation}
At fixed scaled time $\tilde t=t/N$, the large-$N$ limit is
\begin{equation}
    \frac{\langle V_{N\tilde{t}}\rangle}{N\sigma^2}
    \longrightarrow1-e^{-\tilde{t}}.
    \label{eq:supp_variance_scaled}
\end{equation}
The mean within-population variance therefore saturates at
$(N-1)\sigma^2$ on a timescale of order $N$. The population thus has a
finite characteristic width of order $\sigma\sqrt N$ around its center
of mass at long times.

\subsection{Center-of-mass diffusion}

Since $\langle M_{t+1}-M_t\rangle_t=0$, the center of mass is a martingale.
Expanding $M_{t+1}^2$ about $M_t$ eliminates the cross term under the
conditional average:
\[
    \langle M_{t+1}^2\rangle_t
    =M_t^2+\langle(M_{t+1}-M_t)^2\rangle_t.
\]
Averaging over population histories and using
Eq.~\eqref{eq:supp_center_increment_variance} gives
\begin{equation}
    \langle M_{t+1}^2\rangle
    =\langle M_t^2\rangle
    +\frac{\langle V_t\rangle+\sigma^2}{N}.
    \label{eq:supp_center_second_moment_recursion}
\end{equation}
Summation from the monomorphic initial state gives
\begin{align}
    \langle M_t^2\rangle
    &=\sum_{s=0}^{t-1}
        \frac{\langle V_s\rangle+\sigma^2}{N} \notag\\
    &=\sigma^2\left[
        t-(N-1)\left\{1-
        \left(1-\frac{1}{N}\right)^t\right\}
    \right].
    \label{eq:supp_center_msd}
\end{align}
For the scaled center-of-mass position
$\tilde{M}(\tilde{t})=M_{N\tilde{t}}/(\sigma\sqrt{N})$, the same
large-$N$ limit at fixed $\tilde t$ gives
\begin{equation}
    \left\langle\tilde{M}(\tilde{t})^2\right\rangle
    \longrightarrow\tilde{t}-1+e^{-\tilde{t}}.
    \label{eq:supp_center_msd_scaled}
\end{equation}
For $\tilde t\gg1$, the leading behavior is
$\langle\tilde M(\tilde t)^2\rangle\simeq\tilde t$.
The center-of-mass mean-square displacement therefore grows linearly in
time even after the within-population variance has saturated. For
distant first passage, we approximate this long-time motion by Brownian
diffusion. Matching its mean-square displacement to this result fixes
the diffusion coefficient at $1/2$ in scaled units.

\section{Numerical simulations and analysis}
\label{sec:supp_numerics}
\label{sec:supp_numerical_diagnostics}

We compute the scaled mean first-passage time
$\langle\tilde T_L\rangle=\langle T_L\rangle/N$ after running every neutral
realization until first passage.

For each population size, we locate $\tilde L_-$ using the main-text
condition $2\langle\tilde T_L\rangle=\tilde L$. We interpolate
$\log[2\langle\tilde T_L\rangle/\tilde L]$ linearly in $\log\tilde L$ to its
first upward zero crossing. This gives
$A_{\mathrm{pred}}=1/(2\tilde L_-^2)$ and
$\tilde L_+^{\mathrm{pred}}=2\tilde L_-^2$.
We estimate the measured cubic coefficient with a power-law fit whose
exponent is fixed at three in the interior window
$1.2\tilde L_-\leq\tilde L\leq0.8\tilde L_+^{\mathrm{pred}}$.
The fitted coefficient is the geometric mean of
$\langle\tilde T_L\rangle/\tilde L^3$ within this window.

We estimate 95\% percentile bootstrap intervals from 500 replicate-level resamples.
Replicate identifiers are resampled jointly across mutation variances
within a population size, preserving the common random-number pairing.
The fitting window and all estimates are recalculated from each
resampled mean curve.

\begin{figure}[!ht]
    \centering
    \includegraphics[width=0.86\linewidth]{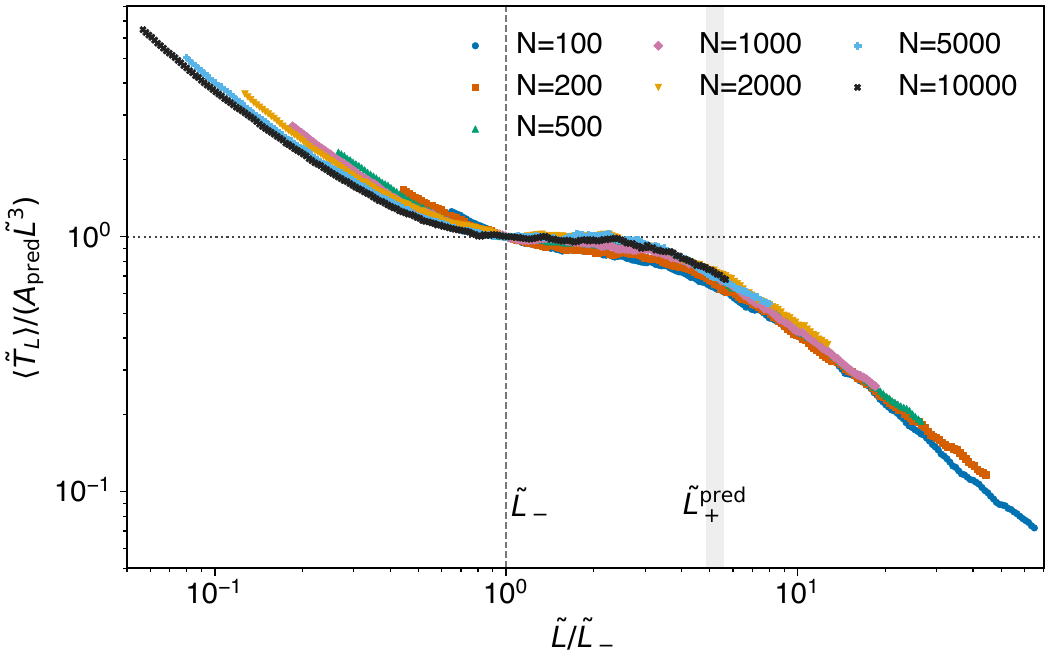}
    \caption{Compensated mean first-passage times.
    The scaled mean first-passage time divided by the predicted cubic form,
    $\langle\tilde T_L\rangle/(A_{\mathrm{pred}}\tilde L^3)$, plotted against
    $\tilde L/\tilde L_-$. Symbols represent means over 1000
    realizations at each parameter point. The horizontal dotted line marks
    unity; the vertical dashed line marks
    $\tilde L/\tilde L_-=1$. The band spans
    $\tilde L_+^{\mathrm{pred}}/\tilde L_-$ across the population-specific
    point predictions, showing variation across $N$, not a confidence
    interval.}
    \label{fig:supp_crossover_validation}
\end{figure}

We compare the fitted cubic coefficient with $A_{\mathrm{pred}}$
inferred from the lower crossover. For $N\geq1000$, the fitted coefficient
is 0.910--0.952 times $A_{\mathrm{pred}}$, so the prediction reproduces
the measured amplitude to within about 10\%. Figure~\ref{fig:supp_crossover_validation}
shows this comparison across target distances by plotting
$\langle\tilde T_L\rangle/(A_{\mathrm{pred}}\tilde L^3)$: values near one throughout
the intermediate regime indicate agreement with the predicted cubic curve.
The interpolated ratio equals one at $\tilde L_-$ by construction; its behavior
beyond this matching point provides the test.

For the same populations, the measured lower crossover
$\tilde L_-=2.703$--$2.801$ gives
$\tilde L_+^{\mathrm{pred}}=14.6$--$15.7$.
This predicted range is consistent with the broad crossover region in
main-text Fig.~2(b), where the measured curves return from cubic to
quadratic behavior.

The gray band in main-text Fig.~2(b) displays the range of predicted
upper-crossover positions, including sampling uncertainty, across all
seven population sizes. It covers their 95\% bootstrap intervals:
the smallest lower bound and largest upper bound give its endpoints,
$11.498$--$16.171$. Its width combines variation across $N$ with
sampling uncertainty in the predicted position; it does not represent
the physical width of the crossover or uncertainty in the matching
approximation.

\clearpage

% TODO: Add supplementary references when manuscript citations are finalized.
% Supplementary citations use the shared reference list above.

\end{document}